%% file: approx.tex
\documentclass[hidelinks]{llncs}

\usepackage{xspace,amsmath}
\usepackage{color}
\usepackage{booktabs}
\usepackage[utf8]{inputenc}
\usepackage{comment}
\usepackage{cite}
\usepackage[hyphens]{url}
\usepackage{graphicx}
\usepackage{siunitx}
\usepackage{mathtools}
\usepackage{amssymb}
\usepackage[T1]{fontenc}
\usepackage{enumitem}

\newcommand{\exclude}[1]{}

\newcommand{\q}{\phantom0}

\newcounter{lnoc}

\newcommand{\lno}[1][0]{{\footnotesize\sffamily 
\ifnum#1=0
\stepcounter{lnoc} 
\ifnum\thelnoc<10
\phantom0%
\fi
\thelnoc
\else
\thelnoc.#1
\fi
}\>}

\usepackage{hyperref}
\hypersetup
{
    pdfauthor={Aleksander Cis{\l}ak and Szymon Grabowski},
    pdftitle={A practical index for approximate dictionary matching with few mismatches},
    pdfkeywords={Hamming distance} {mismatches} {dictionary matching} {text indexing} {split index},
    pdfborder={0},
    urlcolor=black,
    colorlinks=false
}

\begin{document}

\title{A practical index for approximate dictionary matching with few mismatches}

\author{Aleksander Cis{\l}ak$^\dag$ and Szymon Grabowski$^\ddag$}
\institute{$^\dag$ Warsaw University of Technology, Faculty of Mathematics and Information Science,\\
ul.~Koszykowa 75, 00--662 Warsaw, Poland, \email{a.cislak@mini.pw.edu.pl} \\
$^\ddag$ Lodz University of Technology, Institute of Applied Computer Science,\\
  Al.\ Politechniki 11, 90--924 {\L}\'od\'z, Poland, \email{sgrabow@kis.p.lodz.pl}
}

\maketitle

\begin{abstract}
Approximate dictionary matching is a classic string matching problem (checking if a query string occurs in a collection of strings) with applications in, e.g., spellchecking, online catalogs, geolocation, and web searchers.
We present a surprisingly simple solution called a {\em split index}, which is based on the Dirichlet principle, for matching a keyword with few mismatches, and experimentally show that it offers competitive space-time tradeoffs.
Our implementation in the C++ language is focused mostly on data compaction, which is beneficial for the search speed (e.g., by being cache friendly). 
We compare our solution with other algorithms and we show that it performs better for the Hamming distance.
Query times in the order of 1 microsecond were reported for one mismatch for the dictionary size of a few megabytes on a medium-end PC.
We also demonstrate that a basic compression technique consisting in $q$-gram substitution can significantly reduce the index size (up to 50\% of the input text size for the DNA), while still keeping the query time relatively low.

\end{abstract}

\input{intro.tex}
\input{related.tex}
\input{our_alg.tex}

\input{results.tex}
\input{concs.tex}

\clearpage
\section*{Appendix A}
\label{datasets}

The following data sets were used in order to obtain the experimental results:

\begin{itemize}

\item
iamerican --- 0.79\,MB, English, available from Linux packages
\item
foster --- 2.67\,MB, English, available at: \url{http://www.math.sjsu.edu/~foster/dictionary.txt}
\item
iamerican-insane --- 5.8\,MB, English, available from Linux packages
\item
DNA --- 20-mers extracted from the genome of Drosophila melanogaster (available at: \url{http://flybase.org/}), sizes: 6.01\,MB, 135.89\,MB, 262.78\,MB, and 627.80\,MB
\item
A list of common English misspellings --- 44.2\,KB (4,261 words), available at: \url{http://en.wikipedia.org/wiki/Wikipedia:Lists_of_common_misspellings/For_machines}
\end{itemize}

\bibliographystyle{abbrv}
\bibliography{approx}

\end{document}

%% file: intro.tex
\section{Introduction}
Dictionary string matching (keyword matching, matching in dictionaries), defined as the task of checking if a query string occurs in a collection of strings given beforehand, is a classic research topic.
In recent years, increased interest in {\em approximate} dictionary matching 
can be observed, where the query and one of the strings from the dictionary 
may only be similar in a specified sense rather than equal.
Approximate dictionary matching is considered a hard problem, since most useful string similarity measures are non-transitive.
On the other hand, matching with mismatches (i.e.~using a Hamming distance) is also a very desired functionality with applications in, i.a., bioinformatics~\cite{kurtz2001reputer, landau2001algorithm}, biometrics~\cite{davida1998enabling}, cheminformatics~\cite{flower1998properties}, circuit design~\cite{girard1997reduction}, and web crawling~\cite{manku2007detecting}.

As indexes supporting approximate matching tend to grow exponentially 
in $k$, the maximum number of allowed errors, it is also a worthwhile goal 
to design efficient indexes supporting only a small $k$.
In this paper, we focus on the problem of dictionary matching with few mismatches (especially one mismatch).
Formally, for a collection $\mathcal{D} = \{d_1, \ldots, d_m\}$ of $|\mathcal{D}|$ strings (words) $d_i$ of total length $n$ over a given alphabet $\Sigma$ (where $\sigma = |\Sigma|$), $I(\mathcal{D})$ is an approximate dictionary index supporting matching with mismatches, if for any  query pattern $P$ it returns all strings $d_j$ from $\mathcal{D}$  such that $Ham(P, d_j) \leqslant k$ (Hamming distance).
As regards the substrings, they are denoted as $S[i_0, i_1]$ (an inclusive range), and indexes are 0-based.

%% file: related.tex
\section{Related work}
\label{sec:related}

Solutions for approximate dictionary matching can be basically divided into two classes: worst-case space and query time oriented, and heuristical ones. Notable results from the first class include the $k$-errata trie by Cole et al.~\cite{cole2004dictionary} which is based on the suffix tree and the longest common prefix structure.
It can be used in various contexts, including full-text and keyword indexing, as well as wildcard matching.
For the Hamming distance and dictionary matching, it uses $O(n + |\mathcal{D}| \frac{(\log |\mathcal{D}|)^k}{k!})$ space and offers $O(m + \frac{(\log |\mathcal{D}|)^k}{k!} \log \log n + occ)$ query time (this also holds for the edit distance but with larger constants).
This was extended by Tsur~\cite{tsur2010fast} who described a structure similar to the one from Cole et al. with time complexity $O(m + \log \log n + occ)$ (for constant $k$) and $O(n^{1+\varepsilon})$ space for a constant $\varepsilon > 0$.
For full-text searching with the Hamming distance, Gabriele et al.~\cite{gabriele2003indexing} provided an index with average search time $O(m + occ)$ and $O(n \log^l n)$ space (for some $l$).
Another theoretical work describing the algorithm which is similar to our split index was given by Shi and Widmayer~\cite{shi1996approximate}, who obtained $O(n)$ preprocessing time and space complexity and $O(n)$ expected search time if $k$ is bounded by $O(m / \log m)$.
They introduce the notion of home strings for a given $q$-gram, which is the set of strings in $\mathcal{D}$ that contain the $q$-gram in the exact form (the value of $q$ is set to $|P|/(k + 1$).
In the search phase, they partition $P$ into $k + 1$ disjoint $q$-grams and use a candidate inspection order to speed up finding the matches with up to $k$ edit distance errors.

On the practical front, Bocek et al.~\cite{bocek2007fast} provided a generalization of the Mor--Fraenkel~\cite{mor1982hash} algorithm for $k \geqslant 1$ which is called \emph{FastSS}.
To check if two strings $S_1$ and $S_2$ match with up to $k$ errors, we first delete all possible ordered subsets of $k^\prime$ symbols for all $0 \leqslant k^\prime \leqslant k$ from $S_1$ and $S_2$.
Then we conclude that $S_1$ and $S_2$ \emph{may} be in edit distance at most $k$ if and only if the intersection of the resulting lists of strings
is non-empty (explicit verification is still required).
For instance, if $S_1$ = \texttt{abbac} and $k = 2$, then its neighborhood is as follows: \texttt{abbac}, \texttt{bbac}, \texttt{abac}, \texttt{abac}, \texttt{abbc}, \texttt{abba}, \texttt{abb}, \texttt{aba}, \texttt{abc}, \texttt{aba}, \texttt{abc}, \texttt{aac}, \texttt{bba}, \texttt{bbc}, \texttt{bac} and \texttt{bac} (some of the resulting strings are repeated and they may be removed).
If $S_2$ = \texttt{baxcy}, then its respective neighborhood for $k = 2$ will contain, e.g., the string \texttt{bac}, but the following verification will show that
$S_1$ and $S_2$ are in edit distance greater than 2.
If, however, $Lev(S_1 , S_2) \leqslant 2$ (Levenshtein distance), then it is impossible not to have in the neighborhood of $S_2$ at least one string from the
neighborhood of $S_1$, hence we will never miss a match.
The lookup requires $O(k m^k \log(nm^k))$ time (where $m$ is the average dictionary word length) and the index occupies $O(nm^k)$ space.
Another practical filter was presented by Karch et al.~\cite{karch2010improved} and it improved on the FastSS method.
They reduced space requirements and the query time by splitting long words (similarly to FastBlockSS which is a variant of the original method) and storing the neighborhood implicitly with indexes and pointers to original dictionary entries.
They claimed to be faster than other approaches such as the aforementioned FastSS and a BK-tree~\cite{burkhard1973some}.
Recently, Chegrane and Belazzougui~\cite{chegrane2014simple} described another practical index and they reported better results when compared to Karch et al.
Their structure is based on the dictionary by Belazzougui for the edit distance of 1 (see the following subsection).
An approximate (in the mathematical sense) data structure for approximate matching which is based on the Bloom filter was also described~\cite{manber1994algorithm}.

A permuterm index is a keyword index which supports queries with one wildcard symbol~\cite{garfield1976permuterm}.
The idea is store all rotations of a given word appended with the terminating character, for instance for the word \texttt{text}, the index would consist of the following permuterm vocabulary: \texttt{text\$, ext\$t, xt\$te, t\$tex, \$text}.
When it comes to searching, the query is first rotated so that the wildcard appears at the end, and subsequently its prefix is searched for using the index.
This could be for example a trie or any other data structure which supports a prefix lookup.
The main problem with the standard permuterm index is its space usage, as the number of strings inserted into the data structure is the number of words multiplied by the average string length.
Ferragina and Venturini~\cite{ferragina2010compressed} proposed a \emph{compressed} permuterm index in order to overcome the limitations of the original structure with respect to space.
They explored the relation between the permuterm index and the Burrows--Wheeler Transform~\cite{burrows1994block}, which is applied to a concatenation of all strings from the input dictionary.
They provided a modification of the LF-mapping known from FM-indexes~\cite{ferragina2000opportunistic} in order to support the functionality of the permuterm index.

\subsection{The 1-error problem}

It is important to consider methods for detecting a single error, since over 80\% of errors (even up to roughly 95\%) are within $k=1$ for the edit distance with transpositions~\cite{damerau1964technique, pollock1984automatic}.
Belazzougui and Venturini~\cite{belazzougui2012compressed} presented a compressed index whose space is bounded in terms of the $k$-th order empirical entropy of the indexed dictionary.
It can be based either on perfect hashing, having $O(m + occ)$ query time or on a compressed permuterm index with $O(m \min(m, \log_{\sigma} n \log \log n) + occ)$ time (when $\sigma = \log^c n$ for some constant $c$) but improved space requirements.
The former is a compressed variant of a dictionary presented by Belazzougui~\cite{belazzougui2009faster} which is based on neighborhood generation and occupies $O(n \log \sigma)$ space and can answer queries in $O(m)$ time.
Chung et al.~\cite{chung2014efficient} showed a theoretical work where external memory is used, and their focus is on I/O operations.
They limited the number of these operations to $O(1 + m / (wB) + occ / B)$, where $w$ is the size of the machine word and $B$ is the number of words within a block (a basic unit of I/O), 
with the space of the proposed structure of $O(n/B)$ blocks.
In the category of filters, Mor and Fraenkel~\cite{mor1982hash} described a method which is based on the deletion-only 1-neighborhood. 

For the 1-mismatch problem, Yao and Yao~\cite{YaoY95} described the data structure for binary strings with fixed length $m$ with $O(m \log\log |\mathcal{D}|)$ query time and $O(|\mathcal{D}| m \log m)$ space requirements.
This was later improved by Brodal and G\k{a}sieniec~\cite{brodal1996approximate} with a data structure with $O(m)$ query time which occupies $O(n)$ space.
This was in turn extended with a structure with $O(1)$ query time and $O(|\mathcal{D}| \log m)$ space in a cell probe model (where only memory accesses are counted)~\cite{brodal2000improved}.
Another notable example is a recent theoretical work of Chan and Lewenstein~\cite{chan2015fast}, who introduced the index with optimal query time (i.e. $O(m/w + occ)$, where $occ$ is the number of pattern occurrences) which uses additional $O(w d \log^{1+\varepsilon} d)$ bits of space (beyond the dictionary of $d$ strings), assuming a constant-size alphabet.


%% file: our_alg.tex
\section{Our algorithm}
\label{sec:our_alg}

The algorithm that we are going to present is uncomplicated and based on the Dirichlet principle, ubiquitous in approximate string matching techniques.
We partition each word $d$ into $k+1$ disjoint pieces $p_1, \ldots, p_{k+1}$, of average length $|d|/(k+1)$ (hence the name ``split index''), and each such piece acts as a key in a hash table $H_T$.
The size of each piece $p_i$ of word $d$ is determined using the following formula: $|p_i| = \lfloor |d| / (k+1) \rceil$ and $|p_{k+1}| = |d| - \sum_{i = 1}^k |p_i|$, i.e.~the piece size is rounded to the nearest integer and the last piece covers the characters which are not in other pieces.
This means that the pieces might be in fact unequal in length, e.g., 3 and 2 for $|d| = 5$ and $k = 1$.
The values in $H_T$ are the lists of words which have one of their pieces as the corresponding key.
In this way, every word occurs on exactly $k+1$ lists.
This seemingly bloats the space usage, still, in the case of small $k$ the occupied space is acceptable.
Moreover, instead of storing full words on the respective lists, we only store their ``missing'' prefix or suffix.
For instance for the word \texttt{table} and $k=1$, we would have a relation \texttt{tab} $\to$ \texttt{le} on one list (i.e.~\texttt{tab} would be the key and \texttt{le} would be the value) and \texttt{le} $\to$ \texttt{tab} on the other.

In the case of $k=1$, we first populate each list with the pieces without their prefix and then with the pieces without the suffix; 
additionally we store the position on the list (as a 16-bit index) where the latter part begins.
In this way, we traverse only a half of a list on average during the search.
We can also support $k$ larger than 1 --- in this case, we ignore the piece order on a list, and we store $\lceil \log_2(k+1) \rceil$ bits with each piece that indicate which piece of the word (i.e.~where is the missing piece) is the list key.
Let us note that this approach would also work for $k=1$, however, it turned out to be less efficient.

\begin{figure}[h!]
    \centering
    \includegraphics[scale=0.28]{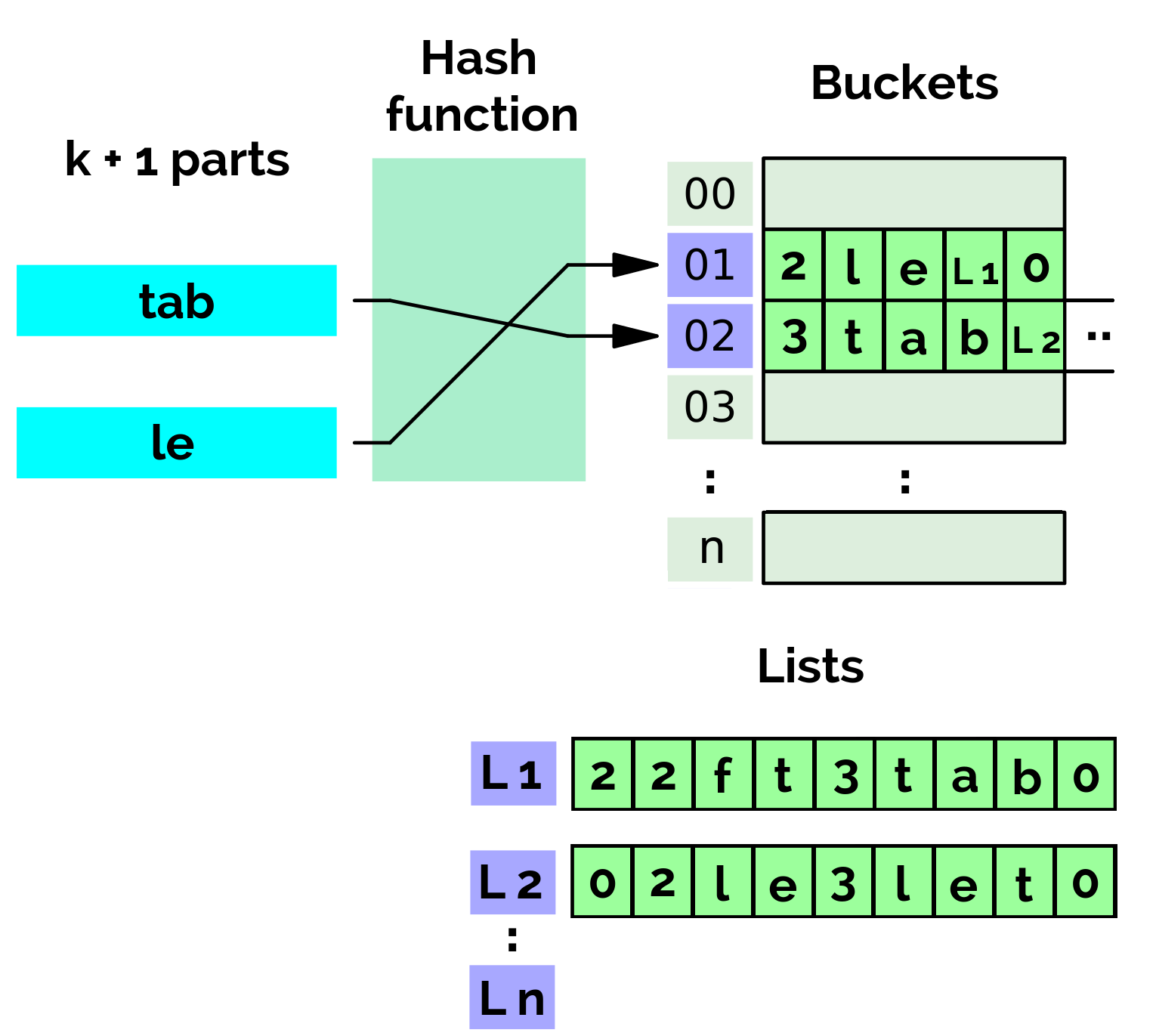}

    \caption[Split index for keyword indexing.]{Split index for keyword indexing which shows the insertion of the word \texttt{table} for $k = 1$. The index also stores the words \texttt{left} and \texttt{tablet} (only selected lists containing pieces of these two words are shown), and \texttt{L1} and \texttt{L2} indicate pointers to the respective lists. The first cell of each list indicates a 1-based word position (i.e.~the word count from the left) where the missing prefixes begin ($k = 1$, hence we deal with two parts, namely prefixes and suffixes), and 0 means that the list has only missing suffixes. Adapted from Wikimedia Commons (author: Jorge Stolfi; available at \url{http://en.wikipedia.org/wiki/File:Hash_table_3_1_1_0_1_0_0_SP.svg}; CC A-SA 3.0).}
    \label{Fig:split_index}
\end{figure}

As regards the implementation, our focus was on data compactness.
In the hash table, we store the buckets which contain word pieces as keys (e.g., \texttt{le}) and pointers to the lists which store the missing pieces of the word  (e.g., \texttt{tab}, \texttt{ft}).
These pointers are always located right next to the keys, which means that unless we are very unlucky, a specific pointer should already be present in the CPU cache during the traversal.
The memory layouts of these substructures are fully contiguous. 
Successive strings are represented by multiple characters with a prepended 8-bit counter which specifies the length, and the counter with the value 0 indicates the end of the list.
During the traversal, each length can be compared with the length of the piece of the pattern.
As mentioned before, the words are partitioned into pieces of fixed length. 
This means that on average we calculate the Hamming distance for only a half of the pieces on the list, since the rest can be ignored based on their length.
Any hash function for strings can be used, and two important considerations are the speed and the number of collisions, since a high number of collisions results in longer buckets, which may in turn have a negative effect on the query time (see Section~\ref{Sec:results} for further discussion).
Figure~\ref{Fig:split_index} illustrates the layout of the split index.

\setcounter{footnote}{0}

The preprocessing stage proceeds as follows:

\begin{enumerate}
\item
Duplicate keywords are removed from the dictionary $\mathcal{D}$.
\end{enumerate}

The following steps refer to each word $d_i$ from $\mathcal{D}$.

\begin{enumerate}[resume]
\item
The word $d_i$ is split into $k + 1$ pieces.
\item
For each piece $p_i$: if $p_i \notin H_T$, we create a new list $L_n$ containing the missing pieces $\mathcal{P} = \{p_j : j \in [1, k + 1] \land j \neq i \}$ and add it to the hash table (we append $p_i$ and the pointer to $L_n$ to the bucket).
Otherwise, if $p_i \in T_H$, we append the missing pieces $\mathcal{P}$ to the already existing list $L_l$.
\end{enumerate}

As regards the search:

\begin{enumerate}
\item
The pattern $P$ is split into $k + 1$ pieces.
\item
We search for each piece $p_i$ (the prefix and the suffix if $k = 1$): the list $L_l$ is retrieved from the hash table or we continue if $p_i \notin H_T$.
Otherwise, we traverse each missing piece $p_j$ from $L_l$.
If $|p_j| = |P| - |p_i|$, the verification is performed and the result is returned if $Ham(p_j, P - p_i) \leqslant k$ (where the subtraction sign indicates substring removal).
\item
The pieces are combined into one word in order to present the answer.
\end{enumerate}

\subsection{Complexity}

Let us consider the average word length $|d|$, where $|d| = (\sum^{|\mathcal{D}|}_{i=1} |d_i|) / |\mathcal{D}|$.
Average time complexity of the preprocessing stage is $O(k n)$, where $k$ is the allowed number of errors, and $n$ is the total input dictionary size (i.e.~the length of the concatenation of all words from $\mathcal{D}$, $n = \sum_{i=1}^{|\mathcal{D}|}|d_i|$).
This is because for each word and for each piece $p_i$ we can either add the missing pieces to a new list or append them to the already existing one in $O(|d|)$ time (if optimized; let us note that $|\mathcal{D}| |d| = n$).
We assume that adding a new element to the bucket takes constant time on average, and that the calculation of all hashes takes $O(n)$ time in total.
This is true irrespective of which list layout is used (there are two layouts for $k = 1$ and $k > 1$, see the preceding paragraphs).
The occupied space is equal to $O(kn)$, because each part appears on exactly $k$ lists and in exactly 1 bucket.

The average search complexity is $O(k t)$, where $t$ is the average length of the list.
We search for each of $k + 1$ pieces of the pattern of length $m$, and when the list corresponding to the piece $p_i$ is found, it is traversed and at most $t$ verifications are performed.
Each verification takes at most $O(\min(m, |d_{max}|))$ time where $d_{max}$ is the longest word in the dictionary\footnote{Or $O(k)$ time, in theory, 
using the old longest common extension (LCE) based technique from Landau and Vishkin~\cite{landau1989fast}, after $O(n\log\sigma)$-time preprocessing.}, but $O(1)$ time on average.
Again, we assume that determining a location of the specific list, that is iterating a bucket, takes $O(1)$ time on average.
As regards the list, its average length $t$ is higher when there is a higher probability that two words $d_1$ and $d_2$ from $\mathcal{D}$ have two parts of the same length $l$
which match exactly, i.e.~$Pr(d_1[i_1, i_1 + l - 1] = d_2[i_2, i_2 + l - 1])$.
Since all words are sampled from the same alphabet $\Sigma$, $t$ depends on the alphabet size, that is $t = f(\sigma)$.
Still, the dependence is rather indirect; in real-world dictionaries which store words from a given language, $t$ will be rather dependent on the $k$-th order entropy of the language.

\subsection{Compression}

In order to reduce storage requirements, we apply a basic compression technique. 
We find the most frequent $q$-grams in the word collection and replace their occurrences on the lists with unused symbols, e.g., byte values $128, \ldots, 255$.
The values of $q$ can be specified at the preprocessing stage, for instance $q = 2$ and $q = 4$ are reasonable for the English alphabet and DNA, respectively.
Different $q$ values can be also combined depending on the distribution of $q$-grams in the input text, i.e.~we may try all possible combinations of $q$-grams up to a certain $q$ value and select ones which provide the best compression.
In such a case, longer $q$-grams should be encoded before shorter ones.
For example, a word \texttt{compression} could be encoded as \texttt{\#p*s\textbackslash} using the following substitution list: $\texttt{com} \to \texttt{\#}, \texttt{re} \to \texttt{*}, \texttt{co} \to \texttt{\$}, \texttt{om} \to \texttt{\&}, \texttt{sion} \to \texttt{\textbackslash}$ (note that not all $q$-grams from the substitution list are used).
Possibly even a recursive approach could be applied, although this would certainly have a substantial impact on the query time.

The space usage could be further reduced by the use of a different character encoding.
For the DNA (assuming 4 symbols only) it would be sufficient to use 2 bits per character, and for the basic English alphabet 5 bits.
In the latter case there are 26 letters, which in a simplified text can be augmented only with a space character, a few punctuation marks, and a capital letter flag.
Such an approach would be also beneficial for space compaction, and it could have a further positive impact on cache usage.
The compression naturally reduces the space while increasing the search time, and a sort of a middle ground can be achieved by deciding which additional information to store in the index.
This can be for instance the length of an encoded (compressed) piece after decoding, which could eliminate some pieces based on their size without performing the decompression and explicit verification.

%% file: results.tex
\section{Experimental results}
\label{Sec:results}

Experimental results were obtained on the machine equipped with the Intel i5-3230M processor running at 2.6\,GHz and 8\,GB DDR3 memory, and the C++ code was compiled with clang version 3.4-1 and run on the Ubuntu 14.04 OS.

One of the crucial components of the split index is a hash function.
Ideally, we would like to minimize the average length of the bucket (let us recall that we use chaining for collision resolution), however, the hash function should be also relatively fast because it has to be calculated for each of the $k + 1$ parts of the pattern (of total length $m$).
We investigated various hash functions, and it turned out that the differences in query times are not negligible, although the average length of the bucket was almost the same in all cases (relative differences were smaller than 1\%).
We can see in Table~\ref{Tab:split_hash} that the fastest function was the xxhash (available on the Internet under the following link: \url{https://code.google.com/p/xxhash/}), and for this reason it was used for the calculation of other results.

\begin{table}[h]
\centering
\begin{tabular}{c|c}
Hash function & Query time (\SI{}{\micro\second}) \\
\hline
xxhash & 0.93\\
sdbm & 0.95\\
FNV1 & 0.95\\
FNV1a & 0.95\\
SuperFast & 0.96\\
Murmur3 & 0.97\\
City & 0.99\\
FARSH & 1.00\\
SpookyV2 & 1.04\\
Farm & 1.04\\
\end{tabular}
\vspace{4mm}
\caption[Evaluated hash functions and search times per query.]{Evaluated hash functions and search times per query for the English dictionary of size 2.67\,MB and $k = 1$. A list of common English misspellings was used as queries, max LF = 2.0.}
\label{Tab:split_hash}
\end{table}

\begin{figure}[h]
    \centering
    \includegraphics[scale=0.55]{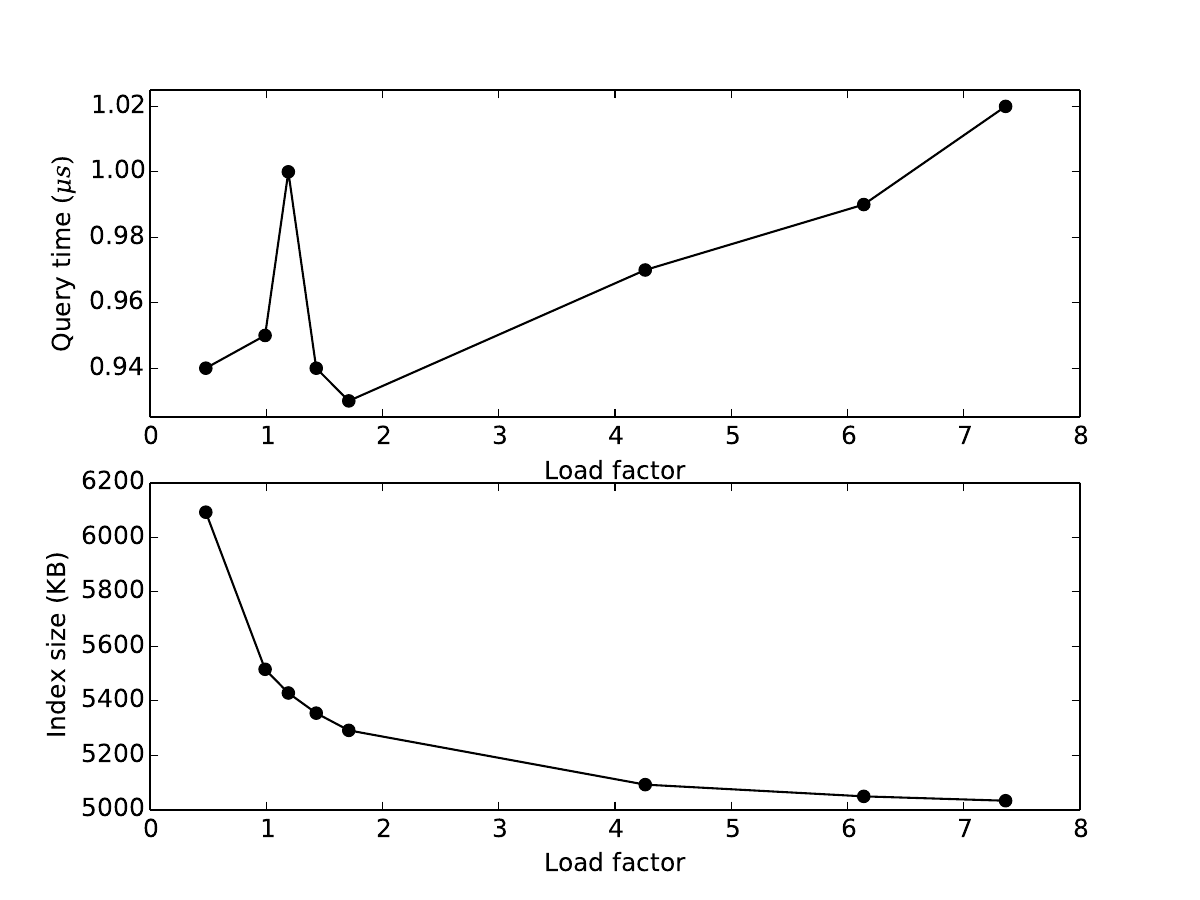}

    \caption[Query time and index size vs the load factor.]{Query time and index size vs the load factor for the English dictionary of size 2.67\,MB and $k = 1$. A list of common English misspellings was used as queries. The value of LF can be higher than 1.0 because we use chaining for collision resolution.}
    \label{Fig:split_lf}
\end{figure}

Decreasing the value of the load factor (LF) did not strictly provide a speedup in terms of the query time, as demonstrated in Figure~\ref{Fig:split_lf}.
This can be explained by the fact that even though the relative reduction in the number of collisions was substantial, the absolute difference was equal to at most a few collisions per list.
Moreover, when the LF was higher, pointers to the lists could be possibly closer to each other, which might have had a positive effect on cache utilization.
The best query time was reported for the maximum LF value of 2.0, 
hence this value was used for the calculation of other results.

In Table~\ref{Tab:split_k} we can see a linear increase in the index size and an exponential increase in query time
with growing $k$.
Even though we concentrate on $k = 1$ and the most promising results are reported for this case, our index might remain competitive also for higher $k$ values.

\begin{table}[h]
\centering
\begin{tabular}{r|cc}
$k$ & Query time (\SI{}{\micro\second}) & Index size (KB) \\
\hline
1 & \q0.51 & 1,715 \\
2 & 11.49 & 2,248 \\
3 & 62.85 & 3,078 \\
\end{tabular}
\vspace{4mm}
\caption[Query time and index size vs the error value $k$.]{Query time and index size vs the error value $k$ for the English language dictionary of size 0.79\,MB. A list of common English misspellings was used as queries.}
\label{Tab:split_k}
\end{table}

$Q$-gram substitution coding provided a reduction in the index size, at the cost of increased query time.
$Q$-grams were generated separately for each dictionary $\mathcal{D}$ as a list of 100 $q$-grams which provided the best compression for $\mathcal{D}$, i.e.~they minimized the size of all encoded words, $S_E = \sum_{i=1}^{|\mathcal{D}|} |Enc(d_i)|$.
For the English language dictionaries, we also considered using only 2-grams or only 3-grams, and for the DNA only 2-grams (a maximum of 25 2-grams) and 4-grams, since mixing the $q$-grams of various sizes has a further negative impact on the query time.
For the DNA, the queries were generated randomly by introducing noise into words sampled from dictionary, and their length was equal to the length of the particular word.
Up to 3 errors were inserted, each with a 50\% probability.
For the English dictionaries we opted for the list of common misspellings, and the results were similar to the case of randomly generated queries.

\begin{figure}[h]
    \centering
    \includegraphics[scale=0.55]{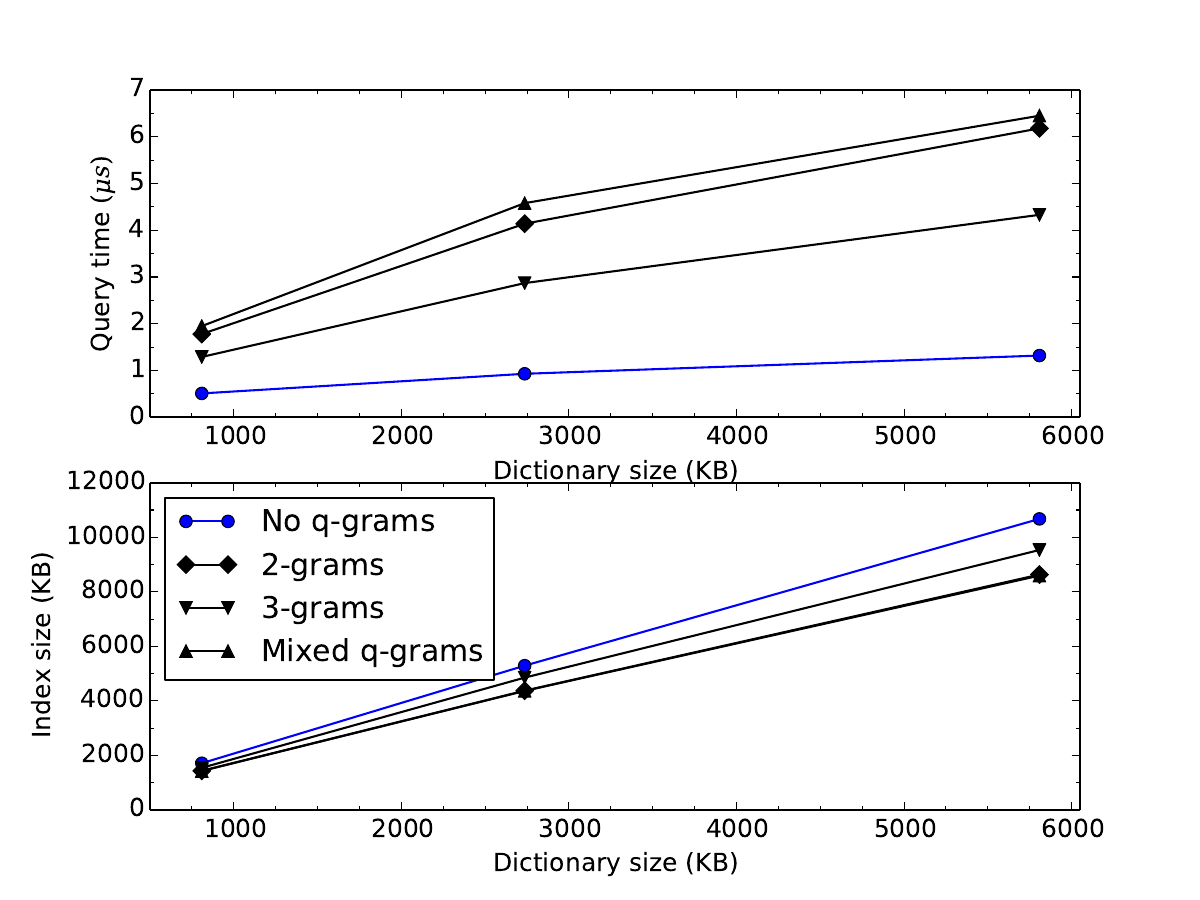}

    \caption[Query time and index size vs dictionary size, with and without $q$-gram coding for English dictionaries.]{Query time and index size vs dictionary size for $k = 1$, with and without $q$-gram coding. Mixed $q$-grams refer to the combination of $q$-grams which provided the best compression, and for the three dictionaries these were equal to ([2-, 3-, 4-] grams): [88, 8, 4], [96, 2, 2], and [94, 4, 2], respectively. English language dictionaries and the list of common English misspellings were used.}
    \label{Fig:split_comp_eng}
\end{figure}

\begin{figure}[h]
    \centering
    \includegraphics[scale=0.55]{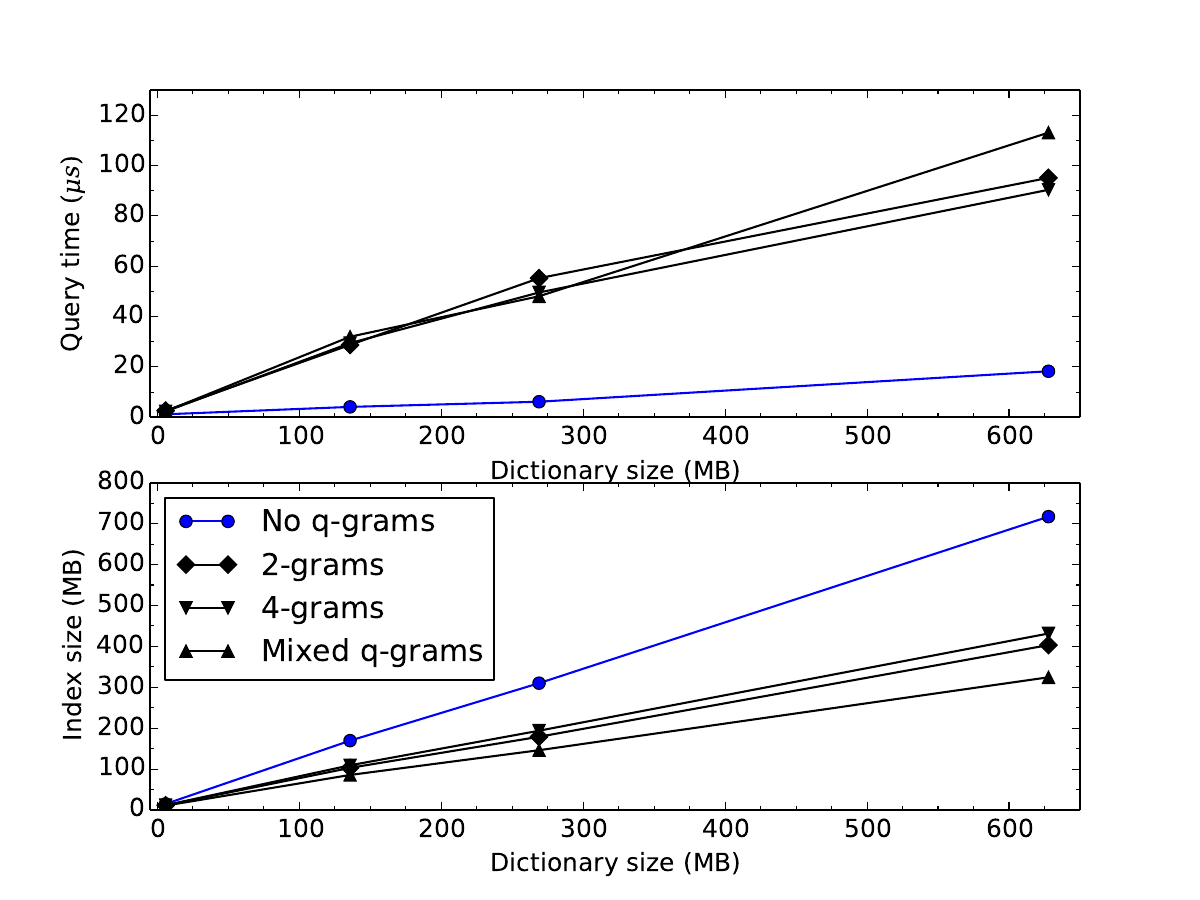}

    \caption[Query time and index size vs dictionary size, with and without $q$-gram coding for DNA dictionaries.]{Query time and index size vs dictionary size for $k = 1$, with and without $q$-gram coding. Mixed $q$-grams refer to the combination of $q$-grams which provided the best compression, and these were equal to ([2-, 3-, 4-] grams): [16, 66, 18] (due to computational constraints, they were calculated only for the first dictionary, but used for all four dictionaries). DNA dictionaries and the randomly generated queries were used.}
    \label{Fig:split_comp_dna}
\end{figure}

We can see the speed-to-space relation for the English dictionaries in Figure~\ref{Fig:split_comp_eng} and for the DNA in Figure~\ref{Fig:split_comp_dna}.
In the case of English, using the optimal (from the compression point of view, i.e.~minimizing the index size) combination of mixed $q$-grams provided almost the same index size as using only 2-grams.
Substitution coding methods performed better for the DNA (where $\sigma = 5$) because the sequences are more repetitive.
Let us note that the compression provided a higher relative decrease in index size with respect to the original text as the size of the dictionary increased.
For instance, for the dictionary of size 627.8\,MB the compression ratio was equal to 1.93 and the query time was still around \SI{100}{\micro\second}.

\begin{figure}[h]
    \centering
    \includegraphics[scale=0.52]{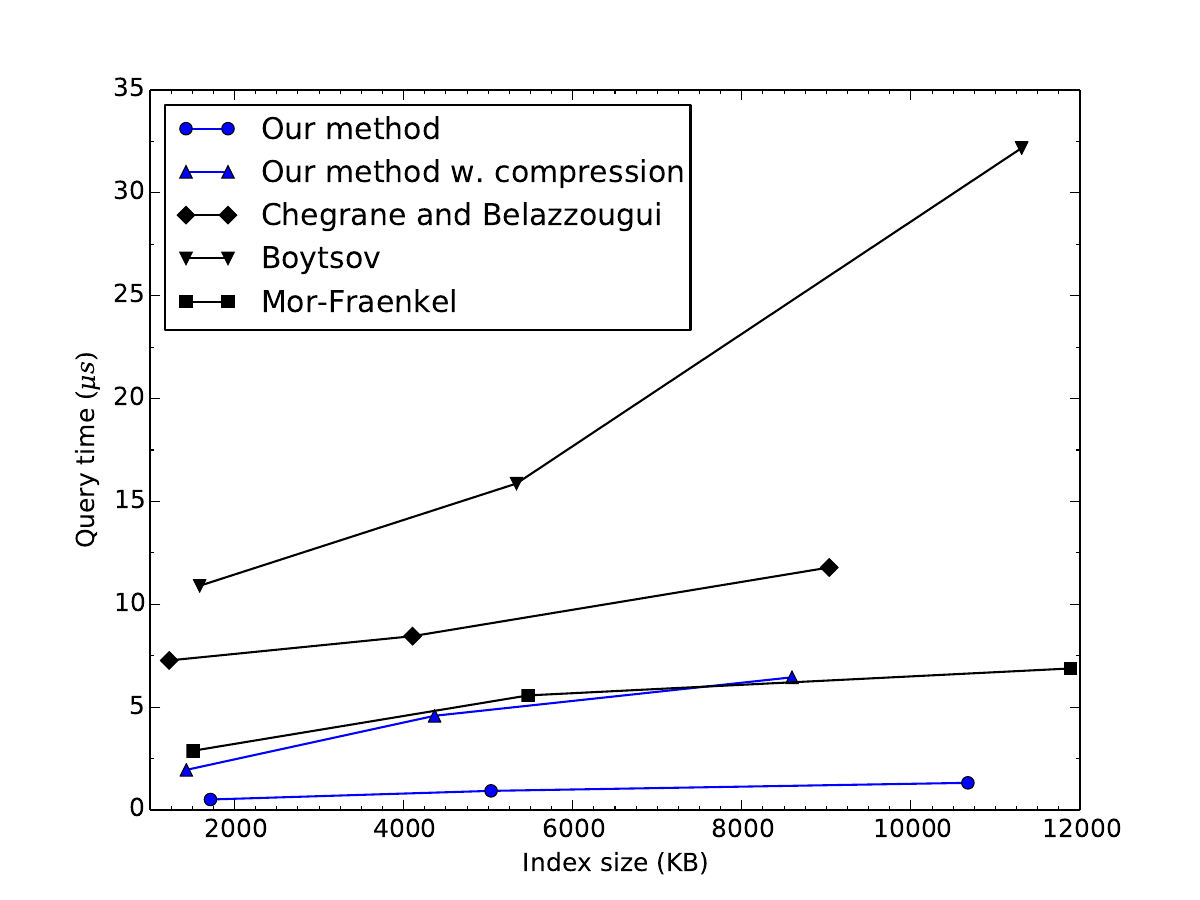}

    \caption[Query time vs index size for different methods.]{Query time vs index size for different methods. The method with compression encoded mixed $q$-grams. We used the Hamming distance, and the other authors used the Levenshtein distance for $k = 1$. English language dictionaries of size 0.79\,MB, 2.67\,MB, and 5.8\,MB were used as input, and the list of common misspellings was used for queries.}
    \label{Fig:split_comp}
\end{figure}

Tested on the English language dictionaries, promising results were reported when compared to methods proposed by other authors.
Others consider the Levenshtein distance as the edit distance, whereas we use the Hamming distance, which puts us at the advantageous position.
Still, the provided speedup is significant, and we believe that the more restrictive Hamming distance is also an important measure of practical use.
The implementations of other authors are available on the Internet (\url{http://searchivarius.org/personal/software};
\url{https://code.google.com/p/compact-approximate-string-dictionary/}, from Boytsov and Chegrane and Belazzougui, respectively).
As regards the results reported for the MF method and Boytsov's Reduced alphabet neighborhood generation, it was not possible to accurately calculate the size of the index (both implementations by Boytsov), and for this reason we used rough ratios based on index sizes reported by Boytsov for similar dictionary sizes.
Let us note that we compare our algorithm with Chegrane and Belazzougui, who report better results when compared to Karch et al., who in turned claimed to be faster than other state-of-the-art methods~\cite{chegrane2014simple, karch2010improved}.
We have not managed to identify any practice-oriented indexes for matching in dictionaries over any fixed alphabet $\Sigma$ dedicated for the Hamming distance, which could be directly compared to our split index.
The times for the brute-force algorithm are not listed, since they were roughly 3 orders of magnitude higher than the ones presented.
Consult Figure~\ref{Fig:split_comp} for details.

We also evaluated different word splitting schemes.
For instance for $k = 1$, one could split the word into two parts of different sizes, e.g., $6 \to (2, 4)$ instead of $6 \to (3, 3)$, however, unequal splitting methods caused slower queries when compared the the regular one.
As regards Hamming distance calculation, it turned out that a naive implementation (i.e.~simply iterating and comparing each character) was the fastest one.
The compiler with automatic optimization was simply more efficient than other implementations (e.g.,~ones based directly on SSE instructions) that we have investigated.

%% file: concs.tex
\section{Conclusions}

We have presented an index for dictionary matching with mismatches, which performed best for the Hamming distance of one.
Its functionality could be extended by storing additional information in the lists that contain the missing parts of the words.
This could be for instance a mapping of words to positions in the document, which would create 
an inverted index supporting approximate matching.

The algorithm can be sped up by means of parallelization, since access to the index during the search procedure is read-only.
In the most straightforward approach we could simply distribute individual words between multiple threads.
A more fine-grained variation would be to concurrently operate on parts of the word after it has been split up (the number of parts depending on the $k$ parameter), or we could even access in parallel lists which contain candidate prefixes and suffixes.
If we had a sufficient amount of threads at our disposal, these approaches could be combined.